# Nonlinear Density Wave Theory and Evolutionary Direction of Spiral Galaxies

Yi-Fang Chang

Department of Physics, Yunnan University, Kunming, 650091, China

(e-mail: yifangchang1030@hotmail.com)

**Abstract**：Based on the nonlinear equations of the density wave theory, the evolutionary direction and the observable conditions on spiral galaxies may be derived by the qualitative analysis theory.
**Key words**： nonlinear equation, spiral galaxy, theory of density wave, evolution

An important problem in the evolution on celestial bodies is of galaxy. According to the classification of galaxies, a major type of them is spiral galaxy. Recently, Pilyugin, et al., discussed the chemical evolution of the Milky Way Galaxy [1]. Perdang, et al., discussed the cellular automation experiments on local galactic structure from model assumptions to numerical simulations [2]. Bianchi, et al., used three-dimensional Monte Carlo method to simulate extinction and polarization of spiral galaxies [3]. Since 1964, C.C.Lin, et al., applied the theory of density to describe better the spiral structure of disk galaxy [4-6]. By using the density wave theory, Li Chi-bin proposed a new approach on the origin of the solar system [7], and Hu Wei-rui discussed the properties of galactic shock waves [8]. Then the density wave theory is applied to study the barred-spiral galaxy, and obtain a unified dynamical explanation for Hubble classification of galaxies [9]. In 1996 the density wave theory is used for to explain gigantic scaling magnetic spiral arms of galaxy NGC6946.

Griv, et al., proposed drift-resonant Landau excitation of spiral density waves in a stellar disk of flat galaxies. This excitation of waves is suggested as a mechanism for the formation of structural features such as spiral arms and the slow dynamical relaxation of galaxies in a regime of hydrodynamical Jeans-type stability [10]. Kondoh, et al., applied nonlinear density wave theory to the spiral structure of galaxies. Their spiral arms are characterized by a soliton and explained as a pattern of a propagating nonlinear density wave [11].

Luo, et al., discussed the spiral magnetohydrodynamic density waves with a tangential shear force and the stability of density waves [12], and the multiwavelength structural manifestations of fast magnetohydrodynamic density waves in spiral galaxies, whose theory provides a basic framework for studying large-scale dynamics and multiwavelength diagnostics of galactic structures [13]. Chi discussed also the resonantly excited nonlinear density waves in disk systems [14]. Cartin, et al., applied the spiral galaxy as a self-regulated system far from equilibrium to look at a reaction-diffusion model for the formation of spiral structures in certain types of galaxies. In numerical runs of the model, spiral structure forms and persists over several revolutions of the disk, but eventually dies out [15]. Viznyuk, et al., investigate the braneworld model and inequality of the spiral galaxy [16]. Saffari, et al., used metric formalism in f(R) modified gravity to study the dynamics of various systems from the solar system to the cosmological scale, and describe the flat rotation curve of the spiral galaxies [17].

In the simulation of the evolutionary process of spiral galaxy, the maintenance of the spiral structures is a big and difficult problem. For this some astronomers proposed various models [18,19]. Further, how is origin of the spiral arms? For multiwavelength observations of spiral galaxies, Luo, et al., emphasized the important perspective that large-scale magnetic fields of MHD density waves play the key role in organizing correlated spiral structures of the various underlying interstellar medium components [13]. The spiral arms of galaxies tend towards tight or pine limit, this problem of evolutionary direction is not still solved completely up to now.

In the density wave theory the equations are [4,5,11]:

$$\sigma_t + [(r\sigma u)_r + (\sigma v)_\vartheta]/r = 0, \qquad (1)$$

$$u_t + uu_r + (v/r)u_\vartheta - v^2/r = \phi_r, \qquad (2)$$
$$v_t + uv_r + (v/r)v_\vartheta + uv/r = \phi_\vartheta/r, \qquad (3)$$
$$\phi_{rr} + \phi_r/r + \phi_{\vartheta\vartheta}/r^2 + \phi_{zz} = -4\pi G\sigma(r,\vartheta)\delta(z), \qquad (4)$$

where the cylindrical system of coordinates is used, u and v are velocities of the fluid of plane polar coordinates $(r,\vartheta)$, $\sigma$ is a surface mass density, and $\phi$ is the negative of the gravitational potential. The first three equations are restricted to the plane z=0. However, because of the complexity of these equations, so far, the discussions for the density wave theory are based mainly on various linearized approaches. Based on the basic equations of a rotating disk of nebula, we use the qualitative analysis theory of nonlinear equations and obtain a nonlinear dynamics model of formation of binary stars [20]. Under certain conditions a pair of singular points results in the course of evolution, which corresponds to the binary stars. Under other conditions these equations give a single central point, which corresponds to a single star. Further, based on the hydrodynamics and hydromagnetics of nebula, we derived quantitatively the formation of binary stars, in which nonlinear interaction and rotation play very crucial roles [21]. Moreover, the Lorenz model may be derived from the equations of the hydrodynamics, whose two wings in the Lorenz model form just the binary stars. While the linear equations form only a single star. Steinitz and Farbiash established the correlation between the spins (rotational velocities) in binaries, and show that the degree of spin correlation is independent of the components separation. Such a result might be related for example to Zhang non-linear model for the formation of binary stars from a nebula [22].

We apply the qualitative analysis theory to the nonlinear equations (2) and (3). Because these equations are similar with the equations of fluid mechanics, Eqs.(2) and (3) can change into

$$du/dt = v^2/r + \phi_r, \qquad (5)$$
$$dv/dt = -uv/r + \phi_\vartheta/r. \qquad (6)$$

Assume that $\phi_r, \phi_\vartheta$ are independent of u and v, then Eqs.(5) and (6) correspond to a characteristic matrix

$$\begin{pmatrix} A & B \\ C & D \end{pmatrix} = \begin{pmatrix} 0 & 2v/r \\ -v/r & -u/r \end{pmatrix}, \qquad (7)$$

whose characteristic equation is

$$\lambda^2 - (A+D)\lambda + (AD - BC) = 0. \qquad (8)$$

Here

$$A + D = -u/r, \quad AD - BC = 2v^2/r^2. \qquad (9)$$

As long as u>0, then A+D<0 and $AD - BC > 0$. In the velocity field, an equilibrium singularity (u,v) is a stable sink. Further, since $\Delta = (u^2 - 8v^2)/r^2$, when $u^2 > 8v^2$, (u,v) is a node; when $u^2 < 8v^2$, (u, v) is a focus [23].

When (u,v) is a node,

$$u = \frac{dr}{dt} = k_1 e^{\lambda t}, v = r\frac{d\theta}{dt} = k_2 e^{\mu t}. \qquad (10)$$

In this case, $\lambda < \mu < 0$. In the polar coordinates, if the integral constants are zero, their positions will be

$$r = \frac{1}{\lambda}k_1 e^{\lambda t}, \theta = \frac{\lambda}{\mu - \lambda}\frac{k_2}{k_1}e^{(\mu-\lambda)t}. \qquad (11)$$

When (u,v) is a focus,

$$u = \frac{dr}{dt} = e^{at}q\cos(bt + \varphi), v = r\frac{d\theta}{dt} = e^{at}q\sin(bt + \varphi). \qquad (12)$$

If the integral constants are zero, their positions will be

$$r = \frac{q}{a^2 + b^2} e^{at}[b\cos(bt + \varphi) + a\sin(bt + \varphi)], \tag{13}$$

$$\theta = (bt + \varphi) - \frac{a}{b}\ln[b\sin(bt + \varphi) + a\cos(bt + \theta)].$$

From (5) and (6) we derive also

$$\frac{du}{dv} = \frac{v^2 + r\phi_r}{-uv + \phi_\theta}. \tag{14}$$

In a simple case, when $\phi_r = \phi_\vartheta = 0$, it is obtained

$$udu + vdv = 0, u^2 + v^2 = C. \tag{15}$$

It shows that the velocities (u=0 and v=0) at the cores of galaxies are a stable central point, and this position is r=C.

For a particular case, if we consider only the evolutionary problem as time, and assume that $u_r = a_1 u$, $u_\vartheta = b_1 u$, $v_r = a_2 v$, $v_\vartheta = b_2 v$ and $\phi_r, \phi_\vartheta$ are independent of u and v, then Eqs.(2) and (3) change into

$$u' = -a_1 u^2 - b_1 uv/r + v^2/r + \phi_r, \tag{16}$$

$$v' = -a_2 uv - b_2 v^2/r - uv/r + \phi_\vartheta/r. \tag{17}$$

For equations (10) and (11), the characteristic matrix is

$$\begin{pmatrix} A & B \\ C & D \end{pmatrix} = \begin{pmatrix} -2a_1 u - (b_1 v/r) & -(b_1 u - 2v)/r \\ -a_2 v - (v/r) & -a_2 u - (2b_2 v + u)/r \end{pmatrix}, \tag{18}$$

Its characteristic equation is still

$$\lambda^2 - (A + D)\lambda + (AD - BC) = 0. \tag{19}$$

Where

$$AD - BC = (2a_1 a_2 + 2a_1/r)u^2 + (4a_1 b_2/r)uv + 2(b_1 b_2 + 1 + a_2 r)v^2/r. \tag{20}$$

$$A + D = -(2a_1 + a_2 + r^{-1})u - (b_1 + 2b_2)r^{-1}v. \tag{21}$$

If $AD - BC > 0$, then A+D<0 and corresponding equilibrium singularity (u,v) is a stable sink, while A+D>0 and corresponding singularity (u,v) is an unstable source.

From $u'=0$ and $v'=0$, we get

$$[a_2^2 r^2 + (ab_1 b_2 - a_1 b_2^2)r + (1 + b_1 b_2)]v^4 - [b_1 \phi_\vartheta + (a_2 b_1 \phi_\vartheta - a_1 b_2 \phi_\vartheta - \phi_r)r - 2a_2 \phi_r r^2 - a_2^2 \phi_r r^3]v^2 - a_1 r \phi_\vartheta^2 = 0, \tag{22}$$

$$u = \frac{\phi_\vartheta - b_2 v^2}{(a_2 r + 1)v}. \tag{23}$$

From these results we know that v and corresponding u have four solutions, it relates possibly with existence of four spiral arms in spiral galaxy.

Most of the disk systems are characterized by spiral structures. In order to explain the existence of spiral arms, a viewpoint is, should consider that the material rotating velocity round the centre of galaxy [24]. According to the qualitative analysis theory, because u and v are velocities, if they satisfy conditions of sink, namely the spiral galaxy tends gradually to stable sink, galaxy oneself also should be stable. Therefore, it can resolve the maintenance problem of the galaxy on long-time. Conversely, if the velocities satisfy conditions of source, the galaxy is unstable also. When the velocities become continuously bigger, then the spiral arms of galaxy should gradually be tighter, and more sparse galaxies are younger. Its limit seems to correspond to ellipsoidal galaxy. In 2003, it is found that the spiral galaxy becomes the ellipsoidal galaxy, which is younger.

When the velocities become continuously smaller, then the spiral arms of galaxy should gradually be looser, and more sparse galaxies are younger. Two cases all make spiral galaxy oneself tend to disappearance.

If $AD-BC<0$, (u, v) will be a saddle point. In this case gravitation and repulsion all appear, the galaxy will evolve to an irregular galaxy. Therefore, for the stability of spiral arms in galaxy the order parameter of most key should be velocity.

The solar velocity is about 250 km/second in the Milky Way Galaxy, and angular velocity is about 360 degree/250,000,000years. Therefore, according to some observable values and the above conclusion, the Galaxy should be stable. Generally, using the qualitative analysis method of the density wave theory, the evolutionary direction of spiral galaxies may be determined. Further, combining the known observational valid radii and densities of spiral galaxies [25], results should be derived. Moreover, the research of the density wave theory in spiral galaxies should is beneficial to understand origins, evolution and stability problem of galaxies. This method can be also further developed and completed, and other properties of the nonlinear equations of the density wave are also discussed.